# Electromagnetic forces for an arbitrary optical trapping of a spherical dielectric

**Antonio A. R. Neves, Adriana Fontes, Liliana de Y. Pozzo, Andre A. de Thomaz, Enver Chillce, Eugenio Rodriguez, Luiz C. Barbosa and Carlos L. Cesar**

*CePOF, Instituto de Física, Universidade Estadual de Campinas, Brazil*
*aneves@ifi.unicamp.br*

**Abstract:** Analytical solution for optical trapping force on a spherical dielectric particle for an arbitrary positioned focused beam is presented in a generalized Lorenz-Mie and vectorial diffraction theory. In this case the exact electromagnetic field is considered in the focal region. A double tweezers setup was employed to perform ultra sensitive force spectroscopy and observe the forces, demonstrating the selectively couple of the transverse electric (TE), transverse magnetic (TM) modes by means of the beam polarization and positioning, and to observe correspondent morphology-dependent resonances (MDR) as a change in the optical force. The theoretical prediction of the theory agrees well with the experimental results. The algorithm presented here can be easily extended to other beam geometries and scattering particles.

## 1. Introduction

One very important contribution of the optical tweezers technique is its ability to extract the missing mechanical measurements in the world of microorganisms and cells that could be correlated to biochemical information. A microsphere displacement is the preferential force transducer for this kind of measurement [1]. Geometrical optics has been used when the particle dimensions are much greater than the light wavelength and Rayleigh scattering theory for the opposite. However, these approximations are no longer valid for the typical conditions used in optical tweezers with very high numerical aperture beams and microspheres with diameters up to ten wavelengths. These conditions require a full vectorial description of the incident beam in partial waves with the origin of coordinate system at the center of the microsphere and not at the focus of the beam. All sorts of approximations and tricks have been used to proceed forward to obtain numerical results. Most of the trapping experiments are performed at the intermediate size regime, between optical geometric and Rayleigh, where diffraction effects are significant. This article presents a theoretical model for different beam configurations, considering diffraction, of a single highly focused beam with arbitrary polarization and focus position with respect to the center of microspheres of any size, together with experimental results to validate the theory.

We have shown that a double optical tweezers can be used to perform an ultrasensitive force spectroscopy by observing forces due to the light scattering and selectively coupling the light to either the TE, the TM or both TE and TM microsphere modes in a single isolated particle as function of the beam polarization and position [6,7]. Our results showed how careful one has to be when using optical force models for mechanical properties measurements. The microsphere modes can change the force values by more than 30-50 %. Also it clearly shows how the usually assumed azimuthal symmetry in the horizontal plane no longer holds because the beam polarization breaks this symmetry. In this article we continue the investigation of the forces in an optical tweezers by using this ultrasensitive experimental technique to measure whole optical force curves as a function of the 3D beam position and polarization. To test optical trapping models it is more important to obtain a full trapping force curve as a function of position instead of just a few points. The observation of diffraction effects and morphology-dependent resonance (MDR) modes (by selectively coupling the light to the transverse electric, transverse magnetic modes in microsphere particles) is a clear demonstration of the validity of our theory.

## 2. Theory

To determine the optical forces rigorously on a dielectric sphere of an arbitrary size, the electromagnetic (EM) equations must be solved for the appropriate boundary conditions. The main effort in the last few years is to have a proper description of the incident EM fields in a highly focused system. This is treated in more detail in a recent publication [2], with a use of a special integral [3], in terms of partial wave expansion coefficients, or beam shape coefficients (BSC), and can be readily applied to Eq. (1). This special integral has been later identified as a special case of the Gegenbauer finite integral. Thus the highly inhomogeneous electric field polarizations generated by a tightly focused arbitrary beam is taken into account by the BSC. The beam focus is generally no longer at the origin of the coordinate system, so all of the beam azimuthal symmetry is lost, this important fact is also taken into account. In this theory the electromagnetic field in an optical trap uses a paraxial Gaussian beam before the objective lens (fulfilling the sine condition) and applying the Angular Spectrum Representation which describes correctly the electromagnetic fields and polarization of a high numerical aperture (NA) objective [2].

$$\mathbf{E}_{inc} = E_0 \sum_{n,m} \left[ \frac{i}{k} G_{nm}^{TM} \nabla \times j_n(kr) \mathbf{X}_{nm}(\theta,\phi) + G_{nm}^{TE} j_n(kr) \mathbf{X}_{nm}(\theta,\phi) \right]$$

$$\mathbf{H}_{inc} = \frac{E_0}{Z} \sum_{n,m} \left[ G_{nm}^{TM} j_n(kr) \mathbf{X}_{nm}(\theta,\phi) - \frac{i}{k} G_{nm}^{TE} \nabla \times j_n(kr) \mathbf{X}_{nm}(\theta,\phi) \right]$$

(1)

Where $\mathbf{X}_{nm}(\theta,\phi) = \mathbf{L} Y / \sqrt{n(n+1)}$ is the vector spherical harmonic, $j_n(kr)$ are spherical Bessel functions and $Z = \sqrt{\mu/\varepsilon}$ is the medium impedance. We only keep the $\sqrt{n(n+1)}$ spherical vectors normalization factor, avoiding plane waves and other $n$ and $m$ dependent factors, to make it easier the verification that each partial wave component has its proper value. The time varying harmonic component $\exp(-i\omega t)$ has been omitted. The EM fields of the scattered wave of the spherical dielectric (of radius $a$) can be written in term of the generalized Mie coefficients as,

$$\mathbf{E}_{sca} = E_0 \sum_{n,m} \left[ \frac{i}{k} a_{nm} \nabla \times h_n^{(1)}(kr) \mathbf{X}_{nm}(\theta,\phi) + b_{nm} h_n^{(1)}(kr) \mathbf{X}_{nm}(\theta,\phi) \right]$$

$$\mathbf{H}_{sca} = \frac{E_0}{Z} \sum_{n,m} \left[ a_{nm} h_n^{(1)}(kr) \mathbf{X}_{nm}(\theta,\phi) - \frac{i}{k} b_{nm} \nabla \times h_n^{(1)}(kr) \mathbf{X}_{nm}(\theta,\phi) \right]$$

(2)

where $h_n^{(1)}(kr)$ is the spherical Hankel function. In a similar way one obtains the fields inside the dielectrical sphere. Satisfying the continuity of the EM fields on the sphere surface, one obtains the generalized Lorenz-Mie coefficients in terms of the parameters that depend on the morphology and incident beam.

$$-a_n = \frac{a_{nm}}{G_{nm}^{TM}} = \frac{M\psi_n(Mx)\psi'_n(x) - \psi'_n(Mx)\psi_n(x)}{\psi'_n(Mx)\xi_n(x) - M\psi_n(Mx)\xi'_n(x)}$$

$$-b_n = \frac{b_{nm}}{G_{nm}^{TE}} = \frac{M\psi'_n(Mx)\psi_n(x) - \psi_n(Mx)\psi'_n(x)}{\psi_n(Mx)\xi'_n(x) - M\psi'_n(Mx)\xi_n(x)}$$

(3)

where $M$ is the relative refractive index, $x$ is the size parameter ($x = ka$), and $\psi$ and $\xi$ are the Riccati-Bessel functions. The Lorenz-Mie coefficients $a_n$ and $b_n$ are responsible for the optical force resonance condition, when the denominator for each TM or TE wave normal modes come close to zero. The BSC, $G_{nm}^{TM}$ and $G_{nm}^{TE}$, are obtained by integrating the radial component of the fields over the solid angle as demonstrated in previous work [2]. Here we use the notation $G_{nm}^{TM}$ instead of the standard $g_{nm}^{TM}$ to emphasize the different normalization factors than that presented by the original authors [5].

To determine the time averaged EM force on a dielectric sphere we proceed by integrating the Minkowski form Maxwell stress tensor over the surface of a sphere in the far field [4]:

$$F_i = \oint T_{ij} n_j dA = \frac{1}{2} \text{Re} \oint [\varepsilon E_i E_j^* + \mu H_i H_j^* - \frac{1}{2}(\varepsilon \mathbf{E} \cdot \mathbf{E}^* + \mu \mathbf{H} \cdot \mathbf{H}^*)\delta_{ij}] n_j dA \quad (4)$$

This can be written as a transversal and longitudinal contribution, where the solid angle integrals are presented in Ref. [5]. Solving them and using the more traditional radiation pressure scattering cross section, we finally obtain:

$$\begin{bmatrix} C_x \\ C_y \end{bmatrix} = \frac{1}{4k^2} \begin{bmatrix} \text{Re} \\ \text{Im} \end{bmatrix} \sum_{n=1} \frac{i}{(n+1)} \left\{ \sqrt{\frac{n(n+2)}{(2n+3)(2n+1)}} \sum_{m=-n}^{n} \sqrt{(n+m+2)(n+m+1)} \right.$$

$$\left[ (a_{n+1} + a_n^* - 2a_{n+1}a_n^*)G_{n+1,-(m+1)}^{TM} G_{n,-m}^{TM*} + (a_n + a_{n+1}^* - 2a_n a_{n+1}^*)G_{nm}^{TM} G_{n+1,m+1}^{TM*} + \right.$$

$$\left. (b_{n+1} + b_n^* - 2b_{n+1}b_n^*)G_{n+1,-(m+1)}^{TE} G_{n,-m}^{TE*} + (b_n + b_{n+1}^* - 2b_n b_{n+1}^*)G_{nm}^{TE} G_{n+1,m+1}^{TE*} \right]$$

$$-\frac{1}{n} \sum_{m=-n}^{n} \sqrt{(n+m+2)(n+m+1)} \sqrt{(n-m)(n+m+1)}$$

$$\left. \left[ (a_n + b_n^* - 2a_n b_n^*)G_{nm}^{TM} G_{n,m+1}^{TE*} - (b_n + a_n^* - 2b_n a_n^*)G_{nm}^{TE} G_{n,m+1}^{TM*} \right] \right\}$$

(5)

$$C_z = \frac{1}{2k^2} \text{Re}[\sum_{n=1} \frac{1}{(n+1)} \left\{ \sqrt{\frac{n(n+2)}{(2n+1)(2n+3)}} \sum_{m=-n}^{n} \left[ \sqrt{(n+m+1)(n-m+1)} \right. \right.$$
$$i(a_n + a_{n+1}^* - 2a_n a_{n+1}^*)G_{nm}^{TM} G_{n+1,m}^{TM*} + i(b_n + b_{n+1}^* - 2b_n b_{n+1}^*)G_{nm}^{TE} G_{n+1,m}^{TE*} \right] \quad (6)$$
$$+ \frac{i}{n} \sum_{m=-n}^{n} m (a_n + b_n^* - 2a_n b_n^*)G_{nm}^{TM} G_{nm}^{TE*} \right\}]$$

The BSCs can be obtained from Ref. [2] for the case of an incident linear $x$-polarized truncated TEM$_{00}$ Gaussian beam, by the expression:

$$\begin{bmatrix} G_{nm}^{TM} \\ G_{nm}^{TE} \end{bmatrix} = \pm 2\pi \, ikf \exp(-ikf) \, i^{n-m} \exp(-im\phi_o) \sqrt{\frac{2n+1}{4\pi n(n+1)} \frac{(n-m)!}{(n+m)!}}$$

$$\int_0^{\alpha_{max}} d\alpha \sqrt{\cos\alpha} \, \exp(-f^2 \sin^2\alpha / \omega_a^2) \exp(-ikz_o \cos\alpha) \quad (7)$$

$$\left\{ \left[ m^2 \frac{J_m(k\rho_o \sin\alpha)}{k\rho_o \sin\alpha} P_n^m(\cos\alpha) - \sin^2\alpha \, J_m'(k\rho_o \sin\alpha) P_n'^m(\cos\alpha) \right] \cos\phi_o \right.$$

$$\left. + im \left[ mJ_m'(k\rho_o \sin\alpha) P_n^m(\cos\alpha) - \sin^2\alpha \frac{J_m(k\rho_o \sin\alpha)}{k\rho_o \sin\alpha} P_n'^m(\cos\alpha) \right] \sin\phi_o \right\}$$

## 3. Materials and methods

We performed the force spectroscopy experiment trapping a 3, 6 and 9 μm polystyrene microsphere diluted in water with a double optical tweezers setup. One beam from a Nd:YAG continuous-wave laser (cw), denoted as a trapping beam, is used to keep the particle trapped, while a second beam from a tunable Ti:Sapphire cw laser, which is modulated (10 Hz) and highly attenuated, denoted as a perturbing beam, is used to perturb the particle from its equilibrium position. Both beams are Gaussian TEM00 laser beam brought to diffraction limited focal spot with a large NA microscope objective (1.25NA 100x oil). We used the same oil immersion objective lens for focusing the trapping beam, the perturbing beam, viewing and collecting the backscattered light. This setup allowed the observation of the optical forces in the axial and radial directions for arbitrary polarization, and the MDR resonances of the TE and TM modes of the microsphere by the selective coupling the beam to any one of the modes. Figure 1 shows a schematic diagram of the optical tweezers configuration used in this work.

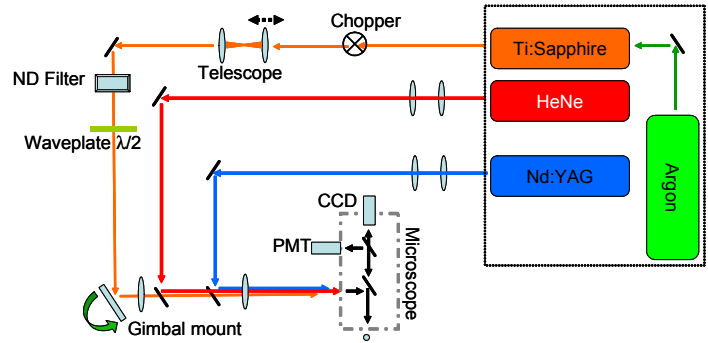

Fig. 1. Complete scheme for the double optical tweezers for ultra-sensitive force spectroscopy.

The Gimbal mount changes the angle of incidence at the back aperture of the objective lens, moving the incident beam focus position laterally (radialy), while the telescope changes the focus depth. A lens system guarantees that the steered beam is pivoted on the objective aperture to avoid power loss. The power transmitted through the objective, measured with an integration sphere as a function of the radial and axial position, was constant. The calibration of the radial beam position was performed by moving a calibrated mark on the Neubauer chamber with the microscope translating stage and using the steering mechanism to positioned the laser beam at the mark, creating a direct correspondence between position and the beam steering step counts. The axial direction calibration procedure is more complicated due to the low axial spatial resolution. For the axial calibration, we attached a half a micron wide fluorescent microsphere to the surface of the Neubauer chamber. With the Ti:Sapphire laser in the femtosecond mode, we moved the microscope stage in the z-direction until a two photon fluorescence maximum intensity was detected with a PMT for several z position step counts of the axial steering mechanism. This way the lateral (radial) and axial steering step counts can be directly translated in terms of position in microns.

To measure the radial optical force on the microsphere as a function of beam position, the microsphere was held in place by the trapping beam while the perturbing beam was moved in the radial direction by using the gimbal mount. The same procedure was adopted for the axial force but moving the first telescope lens instead of the gimbal mount.

A signal proportional to the displacement was measured using the backscattering of a He-Ne laser after passing through two short pass filters to reject the Nd:YAG and Ti:Sapphire laser beams, and detected with a photomultiplier tube (PMT) coupled to the eyepiece of the microscope and a lock-in amplifier. The axial and radial forces were observed by monitoring the amplitude of the displacements while changing the position of the Ti:Sapphire laser. Polarization was controlled by a $\lambda/2$ waveplate.

The MDR were observed by placing the perturbing beam just outside the surface of the sphere where coupling was strongest, with this beam in place the wavelength of the perturbing beam was incremented by rotating the birefringent plate inside the laser cavity with a rotating step motor and calibrated with a monochromator. The fact that the microsphere is suspended in the fluid by the optical tweezers is very important since any near surfaces will ruin the boundary conditions.

## 4. Results and discussion

The microsphere trapped by a highly focused beam will experience a force moving it to a stationary position on the optical axis, generally just behind the focal point, in relation to the propagation direction. This distance is determined by finding the position where the axial force is zero by varying the beam axial position only. Afterwards we simulate the optical forces in the radial direction for different polarization, wavelength and microsphere sizes using the theory presented in Section 2, the numerical convergence is determined by the maximum angular momentum chosen in the expansion, $n_{\max} = k\, r_o$ [2]. Note that all force components are considered since as the beam focuses at the edge, the contribution $z$ force component becomes more significant and reaches the same order of the $x$ and $y$ force components. Here we shall restrict only to the case of radial optical forces, even though this theory applies to optical forces in three dimensions, the axial case has some special considerations that will be addressed in a next publication.

For the radial optical forces, Fig.2, diffraction effects can be seen at the end of figure for either polarizations and reproduced adequately in the theory. Exact fitting with theory was not possible due to unknown exact values of: microsphere size; refractive indexes and beam width at objective aperture. These values affect the diffraction rings transmitted to the optical force. For MDR, Fig. 3 it can be observed that each beam polarization excited each mode, TM or TE, separately as has been demonstrated previously [6]. This can be intuitively understood by noticing that only the radial component of the electric field will excite the TM mode, and only the radial component of the magnetic field will excite the TE modes. It can also be noticed that for the resonant wavelength the radial optical force extends further outside the microsphere (dotted line). MDR resonances enhance optical forces by more than 30-50%.

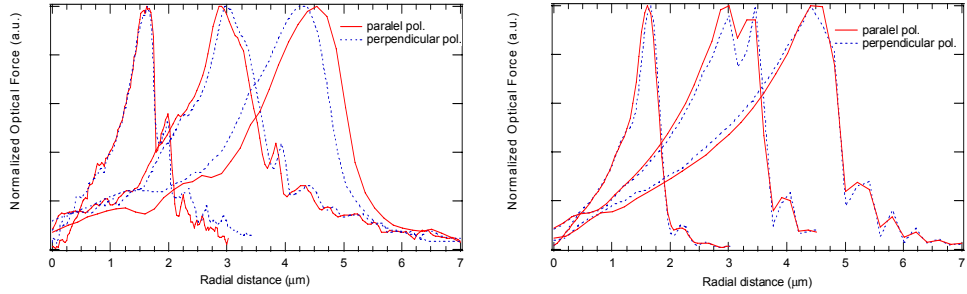

Fig. 2. Radial optical forces for 3, 6 and 9 µm spheres: (left) experimental; (right) theory.

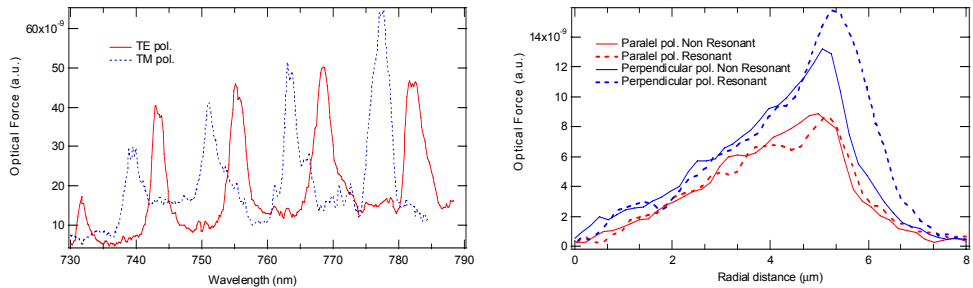

Fig. 3. Radial optical forces enhancement due to MDR for a 9 µm sphere and polarizations as a function of: (left) wavelength; (right) radial distance.

## 5. Conclusions

With this formalism, based on the rigorous solutions of Maxwell equations, proper treatment of optical forces in OT can now be applied taking into account all electromagnetic effects near the trapped dielectric object. We emphasize that the integral (Eq. (3) of Ref. [2]) was fundamental for this development that dramatically simplifies the BSC calculation for an arbitrary translation. We remark that no assumption has been made for the size of the scatterer, thus making it adequate for the most general case of the Mie regime and readily applicable (in the case of optics). The optical forces are described in terms of experimental parameters, depending on the beam profile taking into account the filling factor and dielectric properties of the medium. Not only should a choice of wavelength be considered based on damage to samples, but also at resonance condition as to maximize optical force with wavelength, in this way reducing the power (heating) needed.

The results confirms this theoretical description obtained experimentally also demonstrating how careful one has to be when using optical force models for mechanical properties measurements. The framework of the formalism presented here can also be easily applied to general beams such as Laguerre beams, laser sheets, doughnut and top-hat beams, these types of beams will be deferred to a later study. We have restricted ourselves to the calculation of the radial force since when considering the axial force spherical aberration comes into play and this lengthy discussion is considered for a further publication.

### Acknowledgments

This work was partially supported by Fundação de Amparo à Pesquisa do Estado de São Paulo (FAPESP) through the Optics and Photonics Research Center (CePOF).